\documentclass[journal=jacsat,manuscript=article]{achemso}

\usepackage[version=3]{mhchem}
\usepackage{xcolor}
\usepackage{graphicx}
\usepackage{amsmath}
\usepackage{booktabs}
\usepackage{subcaption}

\title[BeS Ring Resonator Sensor]
{Single-Device VOC Fingerprinting via Polarization-Selective Anisotropic BeS-Clad Microring Resonator}

\author{Sudipta Saha}
\altaffiliation{These authors contributed equally to this work.}

\author{Shoumik Debnath}
\altaffiliation{These authors contributed equally to this work.}

\author{Md Kawsar Alam}
\email{kawsaralam@eee.buet.ac.bd}

\affiliation[BUET]
{Department of Electrical and Electronic Engineering, Bangladesh University of Engineering and Technology, Dhaka 1205, Bangladesh}


\begin{document}

\begin{abstract}
A silicon microring resonator with an anisotropic beryllium sulfide (BeS) cladding is proposed for polarization-selective detection of exhaled-breath biomarkers. The BeS permittivity tensor responds anisotropically to gas adsorption. The transverse-electric (TE) mode probes the in-plane component $\varepsilon_{yy}$ and the transverse-magnetic (TM) mode probes the out-of-plane component $\varepsilon_{zz}$, producing two independent refractive-index readings per analyte from a single device. Five clinically relevant volatile organic compounds are studied: acetone, isoprene, 4-hydroxyhexenal, 2-propenal, and benzene. Anisotropic refractive-index tensors derived from first-principles calculations are used in three-dimensional finite-difference time-domain simulations. The TE mode yields $Q_{TE} = 4520$ and sensitivities of 2.6--5.1~nm/RIU. The TM mode yields $Q_{TM} = 3151$ and sensitivities up to 6.5~nm/RIU. The TE resonance shift is uniform at 0.263~nm across all five analytes and serves as a concentration reference. The TM shift is analyte-specific, ranging from 0.200 to 0.426~nm, and a dispersive TM amplitude inversion uniquely identifies benzene. The pair $(\Delta\lambda_{TE},\,\Delta\lambda_{TM})$ forms a two-dimensional optical fingerprint that distinguishes all five biomarkers. Figures of merit reach $14.9~\text{RIU}^{-1}$ and detection limits reach 1.5~mRIU. Cross-sensitivity simulations for \ce{CO2} and \ce{H2O} confirm that both interferents produce a negative TM resonance shift, placing them in the opposite half-plane from all target biomarkers and providing a natural first-level discriminator. Cladding anisotropy alone provides the chemical selectivity, eliminating the need for a sensor array.
\end{abstract}

\section{Introduction}

Exhaled human breath contains hundreds of volatile organic compounds (VOCs) originating from systemic metabolism, airway chemistry, and environmental exposure~\cite{pauling1971,haick2024,moura2023}. Specific VOCs correlate with disease states, enabling non-invasive diagnostics. Acetone reflects altered fat metabolism and diabetic ketoacidosis~\cite{rydosz2018,ochoa2023}, isoprene tracks cholesterol biosynthesis and respiratory conditions~\cite{king2010}, and aldehydes such as 2-propenal and 4-hydroxyhexenal (4-HHE) indicate oxidative stress and lipid peroxidation~\cite{esterbauer1991}. Benzene, even at low concentrations, is a carcinogenic exposure marker linked to leukemia and lung cancer~\cite{zhang2025}. Despite this promise, simultaneous and sensitive detection of such biomarkers remains challenging for conventional chemiresistive and electrochemical sensors, while gas chromatography--mass spectrometry is limited to laboratory settings~\cite{souzasilva2024,sadeghi2024,nag2023}.

Integrated photonic sensors provide a scalable route to portable breath analysis~\cite{steglich2019,luan2018,hao2023vocml}. Among these, silicon microring resonators (MRRs) offer compact footprints, high quality factors, and CMOS compatibility~\cite{vahala2003,bogaerts2012}. Label-free refractometric sensing has been widely demonstrated for biomolecules and gases using functionalized claddings~\cite{bryan2023,chen2024,shi2022,steglich2019,barrios2007,luan2018,yuan2022mof}. High bulk sensitivities have also been reported in slot-waveguide geometries, reaching tens to hundreds of nm/RIU~\cite{barrios2007,shi2022}. However, a fundamental limitation persists: a single resonator provides only one scalar observable, the resonance shift $\Delta\lambda$. As a result, analytes inducing similar refractive-index perturbations produce indistinguishable responses. Existing approaches --- sensor arrays, selective functionalization, or Vernier configurations --- either increase system complexity or fail to provide truly independent observables~\cite{chen2024,hao2023vocml,yuan2022mof,borga2022}.

This work addresses this limitation by exploiting polarization as an additional sensing dimension. In isotropic claddings, TE and TM responses remain proportional and do not yield independent information~\cite{li2020anisotropic}. This proportionality is broken when the cladding is optically anisotropic and exhibits anisotropic adsorption-induced perturbations. Under these conditions, the TE and TM modes probe orthogonal components of the permittivity tensor, generating two independent observables $(\Delta\lambda_{TE}, \Delta\lambda_{TM})$. The ratio between them becomes analyte-specific, enabling intrinsic chemical discrimination within a single device. While artificially anisotropic media such as plasmonic metamaterials have been explored for enhanced sensing~\cite{kabashin2009}, intrinsic cladding anisotropy has not been widely exploited for VOC discrimination in the telecom regime.

We implement this concept using beryllium sulfide (BeS) as the sensing cladding. BeS is a wide-bandgap II--VI chalcogenide that is transparent in the telecom band and exhibits strong intrinsic optical anisotropy, with a refractive-index contrast of approximately 0.11 between in-plane and out-of-plane components at 1550~nm~\cite{ayirizia2021bes,kumar2022bechalco}. Its surface supports physisorption of polar VOC molecules, and conformal thin films can be realized via atomic layer deposition~\cite{george2010ald}. Adsorption induces anisotropic perturbations in its dielectric tensor, providing the physical basis for polarization-selective sensing.

The contributions of this work are threefold. First, the anisotropic refractive-index response of BeS to five clinically relevant breath biomarkers is obtained from first-principles calculations. Second, a BeS-clad silicon MRR is analyzed using three-dimensional finite-difference time-domain simulations, demonstrating that TE and TM modes probe orthogonal tensor components and yield a two-dimensional optical fingerprint. Third, this fingerprint uniquely distinguishes all analytes, with a dispersive TM amplitude response further isolating benzene, and with cross-sensitivity simulations confirming that dominant breath interferents (\ce{CO2} and \ce{H2O}) fall in a distinct region of the fingerprint plane. The proposed approach enables single-device, array-free chemical discrimination among refractometrically similar VOCs.

\begin{figure}[t]
    \centering
    \includegraphics[width=0.9\linewidth]{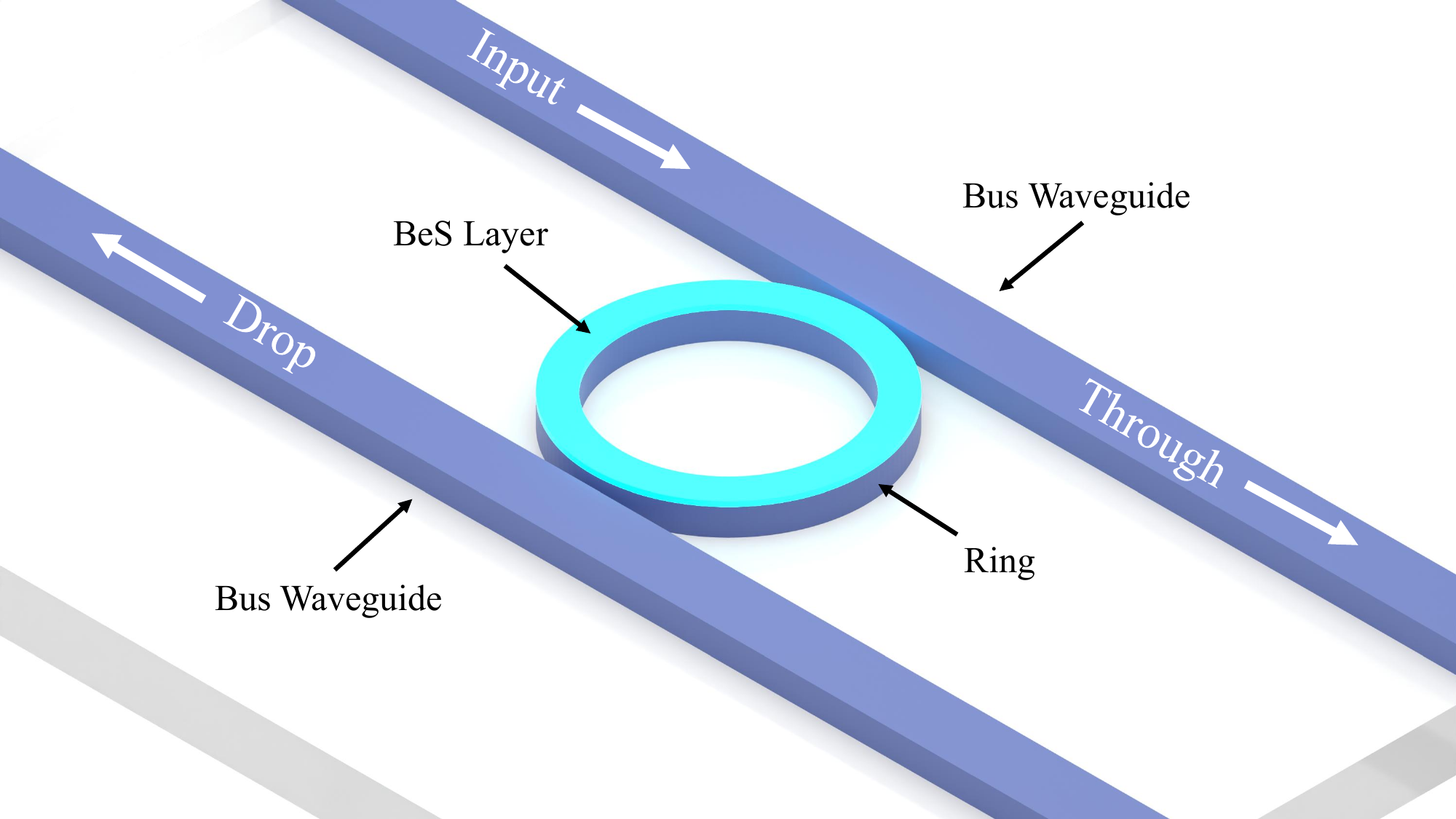}
    \caption{Schematic of the proposed BeS-coated silicon microring resonator sensor. Light enters the input bus, couples into the ring, and exits through the through and drop ports. A conformal BeS film on the ring acts as the anisotropic sensing cladding. Target VOCs adsorb on the BeS surface and perturb its permittivity tensor.}
    \label{fig:schematic}
\end{figure}

\section{Device Design and Operating Principle}

\subsection{Ring Resonator Configuration}
\begin{figure}[t]
    \centering
    \includegraphics[width=0.6\linewidth]{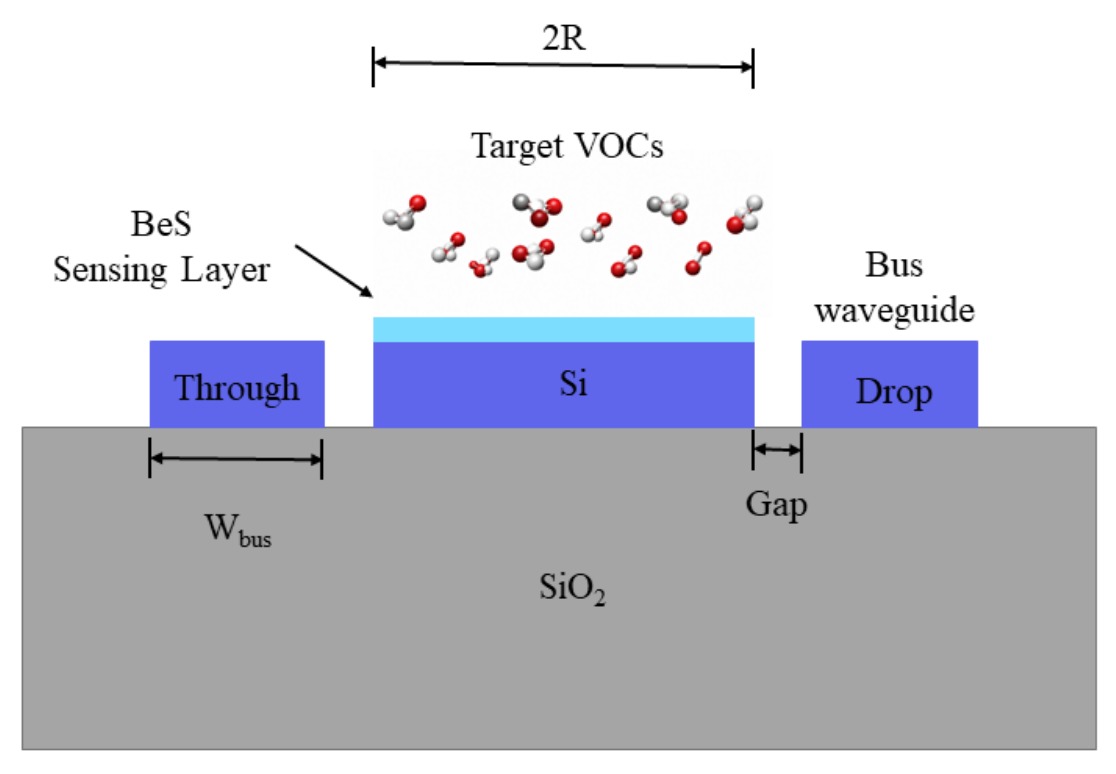}
    \caption{Cross section of the proposed BeS-coated silicon microring resonator sensor.}
    \label{fig:crosssec}
\end{figure}

Figure~\ref{fig:schematic} shows the proposed sensor. A silicon microring is side-coupled to two bus waveguides. The upper bus carries the input and through-port signals. The lower bus carries the drop-port signal. All waveguides are etched in the 220-nm-thick silicon device layer of a silicon-on-insulator (SOI) wafer with a buried oxide lower cladding. The waveguide core is 500~nm wide and 220~nm high, corresponding to a standard single-mode SOI geometry~\cite{selvaraja2009}. The ring radius is 10~$\mu$m, placing a resonance near 1547~nm inside the telecom C-band. The top and sidewalls of the ring are conformally coated with a 10-nm BeS film, which is compatible with atomic layer deposition. The space above the BeS layer is the gas-sensing region. Figure~\ref{fig:crosssec} shows the cross section of the proposed sensor.

\subsection{Operational Baseline and Humidity Equilibration}

In clinical deployment the sensor is never interrogated in vacuum.
Exhaled breath is saturated with water vapor at body temperature,
and the BeS surface equilibrates with ambient \ce{H2O} before any
target biomarker arrives at the ring.
Using the pristine BeS optical constants as the sensing baseline
therefore overestimates the effective cladding contrast and
underestimates the interferent contribution.

A \emph{wet baseline} is therefore defined: the reference state is
taken as BeS with a stable \ce{H2O} monolayer physisorbed at its
dominant surface site rather than bare BeS in vacuum.
First-principles calculations for the \ce{H2O}-adsorbed state show
that water physisorption induces only a 0.04\% change in the BeS
bandgap~\cite{nanoselect2024}, confirming that the optical constants
$n_y$ and $n_z$ are negligibly perturbed from their pristine values
at 1550~nm.
The \ce{H2O}-adsorbed state therefore serves as a well-defined,
physically realistic reference without materially altering the
intrinsic anisotropy that drives the sensing mechanism.
All resonance shifts reported in Section~\ref{sec:results} are
referenced to this wet baseline unless stated otherwise.

\subsection{Polarization Axes and the Permittivity Tensor}
\begin{figure}[htbp]
    \centering
    \includegraphics[width=\linewidth]{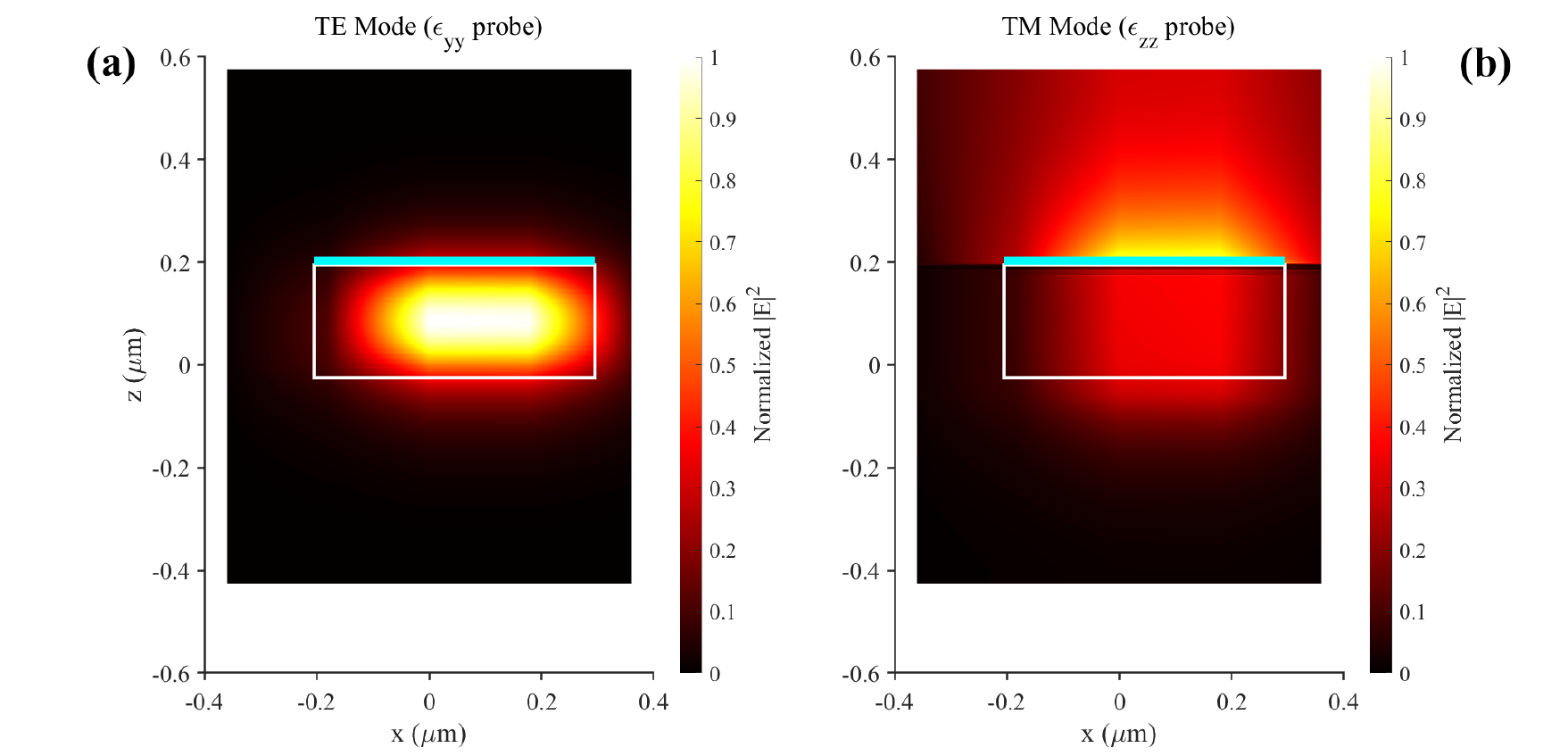}
    \caption{Normalized electric-field intensity distributions of the (a) TE mode and (b) TM mode in the cross-section of the BeS-clad silicon waveguide. The white rectangle outlines the Si core. The cyan strip on top is the 10-nm BeS film. The TE mode concentrates energy in the core and probes the in-plane BeS permittivity $\varepsilon_{yy}$. The TM mode has a pronounced vertical evanescent tail through the BeS film and probes $\varepsilon_{zz}$.}
    \label{fig:modes}
\end{figure}

The device coordinate system follows the simulation setup. The waveguide propagates along the $x$-axis. The waveguide width is along the $y$-axis, and the height is along the $z$-axis. The two fundamental guided modes of the ring are shown in Figure~\ref{fig:modes}. The TE mode has its dominant electric field polarized along $y$, parallel to the BeS film plane. The TM mode has its dominant electric field polarized along $z$, perpendicular to the film. The TE mode therefore couples to the in-plane permittivity component $\varepsilon_{yy}$ of the BeS cladding. The TM mode couples to the out-of-plane component $\varepsilon_{zz}$.

When a VOC molecule adsorbs on the BeS surface, charge redistribution changes the polarizability of BeS bonds along different crystallographic axes by different amounts. The induced change in the permittivity tensor, $\Delta\boldsymbol{\varepsilon}$, has non-equal diagonal entries. The asymmetry between $\Delta\varepsilon_{yy}$ and $\Delta\varepsilon_{zz}$ is the physical origin of the polarization-selective response. A scalar-refractometric sensor would see only a single effective cladding index change. The dual-polarization readout separates this into two independent projections of the tensor.

\subsection{Choice of BeS as Sensing Layer}
BeS was selected for three reasons. First, its bandgap exceeds 5~eV, making it optically transparent across the telecom band and preventing material absorption from degrading the cavity Q-factor~\cite{ayirizia2021bes}. Second, its zinc-blende crystal structure exposes surface sites that favor physisorption of polar carbonyl and unsaturated hydrocarbon species. Third, its strong ionicity produces a measurable and anisotropic change in refractive index upon adsorption, which is the quantity of interest for refractometric sensing. These properties are verified in the next section using first-principles optical constants.

\section{Optical Modeling and Material Properties}

\subsection{First-Principles Refractive Index Tensor}
The diagonal components of the BeS refractive-index tensor were obtained from density functional theory (DFT) calculations using the projector-augmented-wave method~\cite{kresse1996}. Exchange-correlation effects were treated within the generalized gradient approximation. The optical response was computed by evaluating the frequency-dependent dielectric tensor, from which the real and imaginary parts of the refractive index along $x$, $y$, and $z$ were extracted. Calculations were performed for pristine BeS and for BeS with each of the five target biomarkers physisorbed at the dominant surface site. Results were tabulated in Lumerical-compatible $n,k$ files on a 5~nm wavelength step from 400 to 2000~nm.

\subsection{Pristine Anisotropy and Transparency}
\begin{figure}[htbp]
    \centering
    \includegraphics[width=\linewidth]{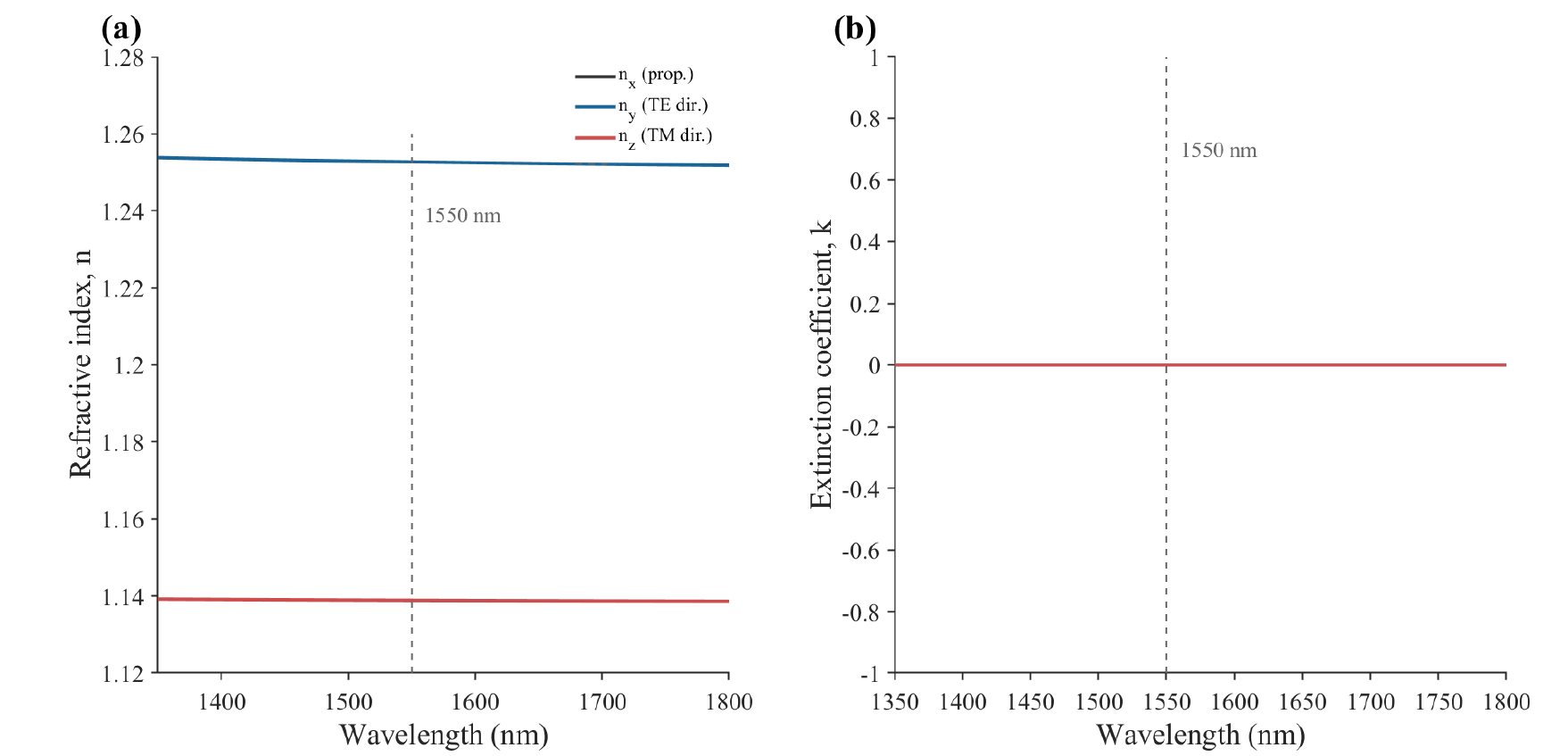}
    \caption{Optical constants of pristine BeS from first-principles calculations. (a) Diagonal components of the refractive-index tensor. $n_x$ (propagation) and $n_y$ (TE direction) are nearly degenerate near 1.253, while $n_z$ (TM direction) is close to 1.139. (b) Extinction coefficient. BeS is effectively transparent across the telecom band.}
    \label{fig:pristine_nk}
\end{figure}

Figure~\ref{fig:pristine_nk} shows the pristine BeS optical constants between 1350 and 1800~nm. The propagation-direction index $n_x$ and the in-plane index $n_y$ both lie close to 1.253 with negligible dispersion. The out-of-plane index $n_z$ is close to 1.139. The two orthogonal projections therefore differ by approximately 0.11, which confirms a strong intrinsic anisotropy of the BeS film~\cite{kumar2022bechalco}. The extinction coefficient in panel (b) is below $10^{-4}$ across the full band. Material absorption does not contribute measurably to the cavity loss, so the Q-factor is set by waveguide scattering and bending loss rather than by the sensing layer.

\subsection{Gas-Induced Index Response}
\begin{figure}[htbp]
    \centering
    \includegraphics[width=\linewidth]{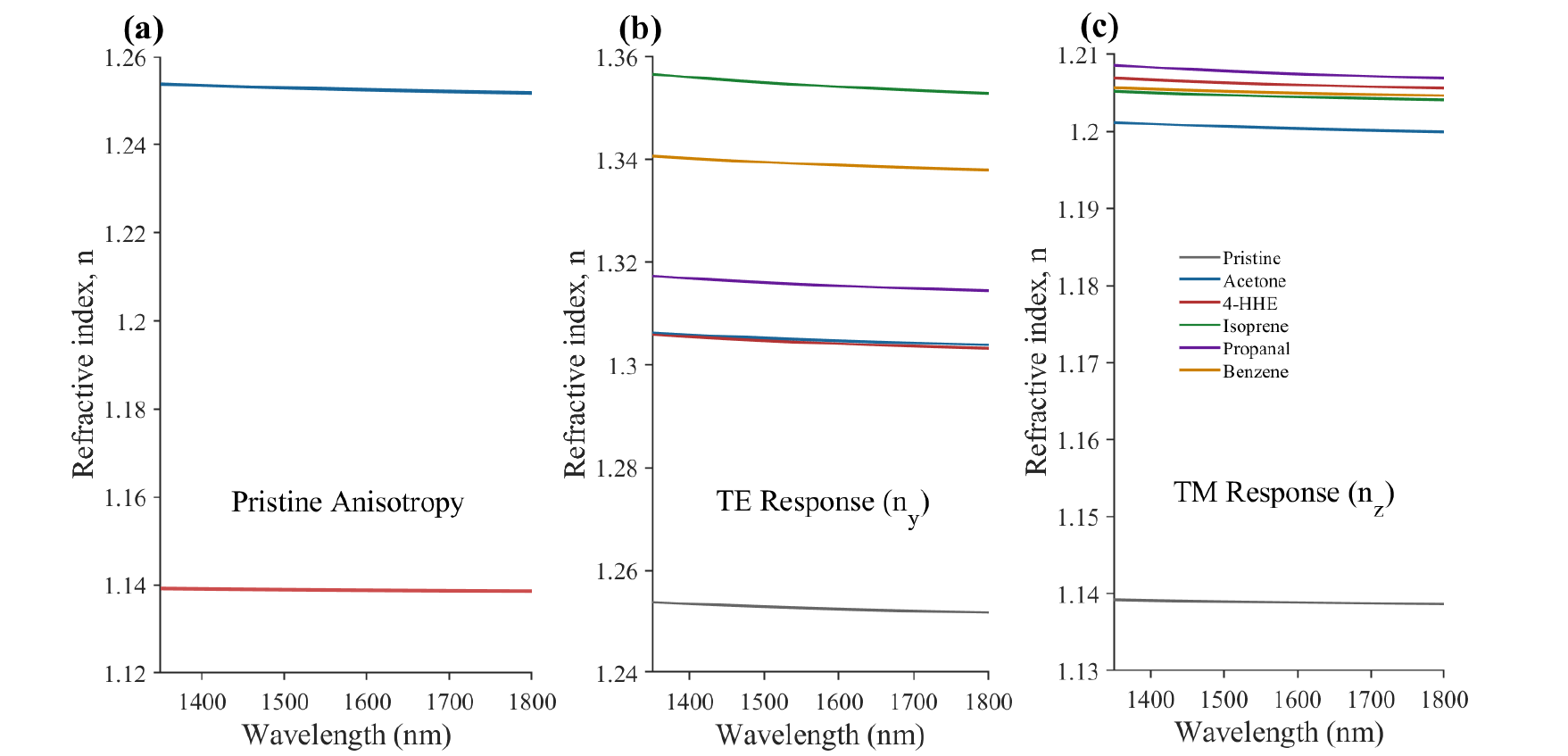}
    \caption{Refractive-index response of BeS to the five target biomarkers. (a) Pristine anisotropy between $n_y$ (upper curve) and $n_z$ (lower curve). (b) TE-probed component $n_y$ after gas adsorption. (c) TM-probed component $n_z$ after gas adsorption. The TE-axis response spans a wider range than the TM-axis response.}
    \label{fig:anisotropy}
\end{figure}

Figure~\ref{fig:anisotropy} summarizes the refractive-index response of BeS to the five biomarkers. Panel (a) restates the pristine anisotropy. Panels (b) and (c) show the TE-probed component $n_y$ and the TM-probed component $n_z$ after adsorption. Both channels shift upward from the pristine baseline for every gas. The magnitudes differ between channels and between analytes. Along the TE axis, isoprene produces the largest shift with $\Delta n_y = 0.102$, and 4-hydroxyhexenal produces the smallest with $\Delta n_y = 0.052$. Along the TM axis, the response falls in a narrower range between $\Delta n_z = 0.062$ for acetone and $\Delta n_z = 0.069$ for 2-propenal.

\begin{figure*}[htbp]
    \centering
    \begin{subfigure}{0.45\textwidth}
        \centering
        \includegraphics[width=\linewidth]{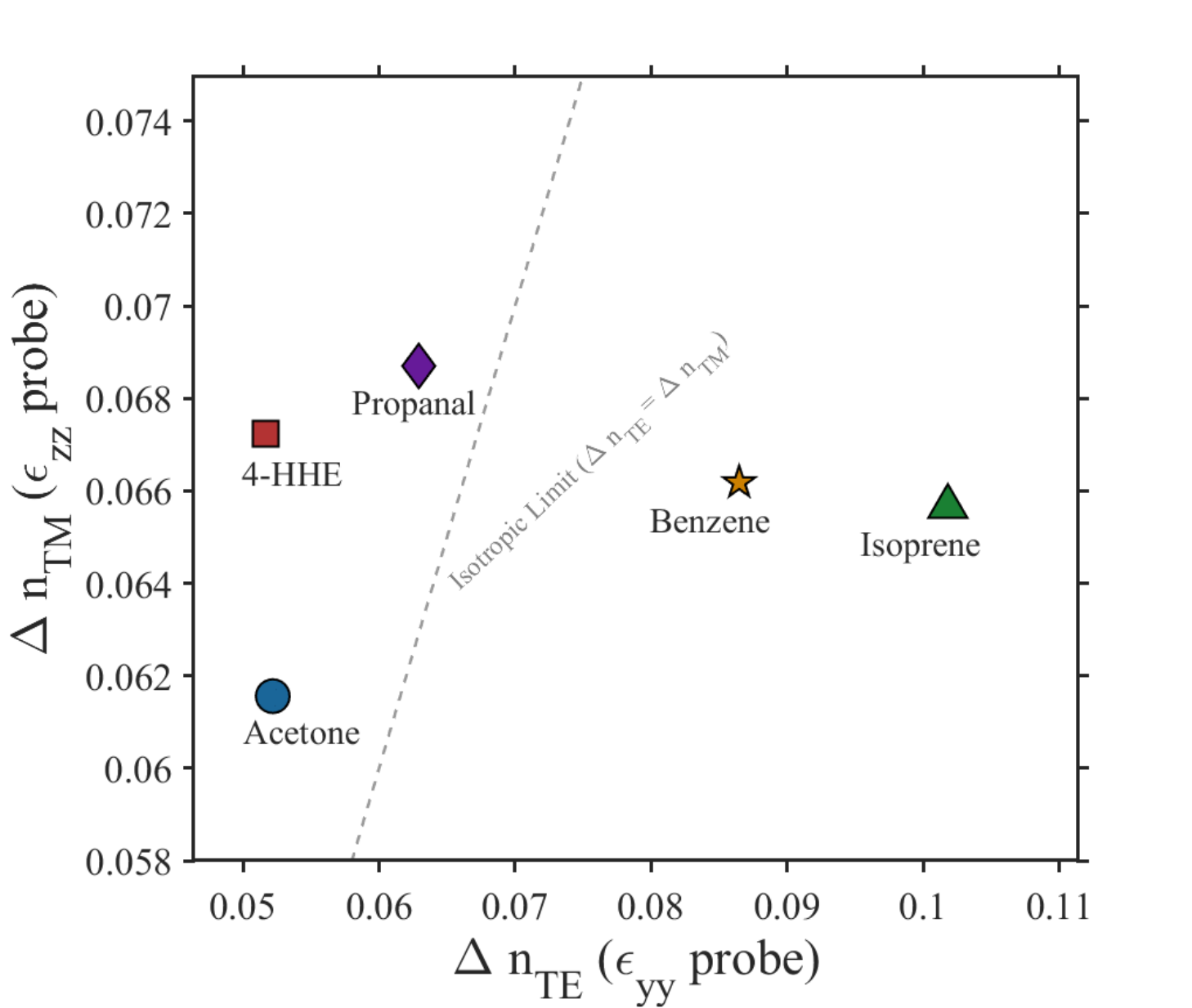}
        \caption{}
        \label{fig:material_fp_a}
    \end{subfigure}
    \hfill
    \begin{subfigure}{0.49\textwidth}
        \centering
        \includegraphics[width=\linewidth]{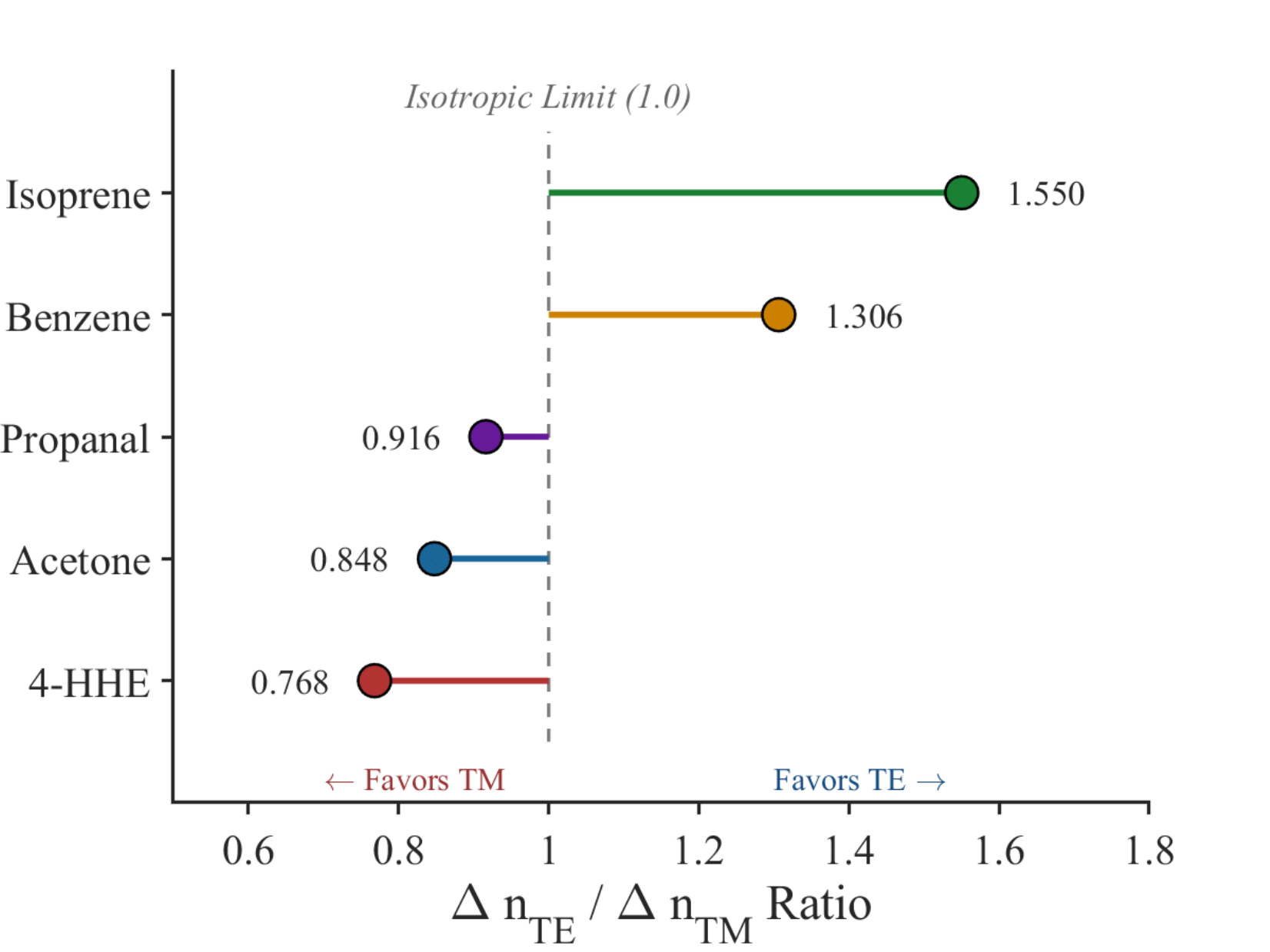}
        \caption{}
        \label{fig:material_fp_b}
    \end{subfigure}
    \caption{Material-level anisotropy fingerprint of BeS.
    (a) $\Delta n_{TM}$ versus $\Delta n_{TE}$ for the five biomarkers. The dashed line marks the isotropic limit. All analytes deviate from this line and occupy distinct coordinates.
    (b) Anisotropy ratio $\Delta n_{TE}/\Delta n_{TM}$. Values above one favor the TE channel and values below one favor the TM channel.}
    \label{fig:material_fp}
\end{figure*}

Figure~\ref{fig:material_fp}(a) plots $\Delta n_{TM}$ against $\Delta n_{TE}$ for each analyte at the operating wavelength. The dashed line is the isotropic limit $\Delta n_{TE} = \Delta n_{TM}$. All five points lie off this line and occupy distinct positions in the plane. No two analytes share the same coordinate. The anisotropy ratio $\Delta n_{TE} / \Delta n_{TM}$ is plotted in Figure~\ref{fig:material_fp}(b). Isoprene and benzene have ratios above unity and favor the TE channel. 4-hydroxyhexenal, acetone, and 2-propenal have ratios below unity and favor the TM channel. The ratio ranges from 0.77 for 4-hydroxyhexenal to 1.55 for isoprene, a factor of two between the extremes. The directional selectivity of the BeS response is therefore both measurable and chemically specific.

\section{Sensing Mechanism and Theoretical Analysis}

\subsection{Resonance Shift from Cladding Perturbation}
The resonance condition of the ring is
\begin{equation}
    m \lambda_{\text{res}} = n_{\text{eff}}(\lambda_{\text{res}})\, L,
    \label{eq:resonance}
\end{equation}
where $m$ is the azimuthal mode number, $n_{\text{eff}}$ is the effective index of the guided mode, and $L = 2\pi R$ is the ring circumference. A perturbation of the cladding refractive index $\Delta n_{\text{clad}}$ shifts the effective index by $\Delta n_{\text{eff}} = \Gamma \, \Delta n_{\text{clad}}$, where $\Gamma$ is the modal confinement factor in the cladding region. Linearizing Equation~\eqref{eq:resonance} gives
\begin{equation}
    \frac{\Delta \lambda}{\lambda_{\text{res}}} = \frac{\Gamma \, \Delta n_{\text{clad}}}{n_g},
    \label{eq:shift}
\end{equation}
where $n_g$ is the group index of the mode. The waveguide sensitivity in nm/RIU is
\begin{equation}
    S = \frac{d\lambda}{d n_{\text{clad}}} = \frac{\lambda_{\text{res}} \, \Gamma}{n_g}.
    \label{eq:sensitivity}
\end{equation}

\subsection{Polarization-Resolved Response}
In the present device $\Gamma$ and $n_g$ differ between TE and TM modes, and the cladding index seen by each mode corresponds to a different component of the BeS tensor. The TE mode responds to $\Delta n_y$ and the TM mode responds to $\Delta n_z$. The two observables are
\begin{align}
    \Delta \lambda_{TE} &= S_{TE} \cdot \Delta n_y(\text{gas}), \label{eq:te_shift}\\
    \Delta \lambda_{TM} &= S_{TM} \cdot \Delta n_z(\text{gas}). \label{eq:tm_shift}
\end{align}
Each gas therefore produces a unique pair $(\Delta \lambda_{TE}, \Delta \lambda_{TM})$ set by the device-level sensitivities $S_{TE}$, $S_{TM}$ and the gas-specific anisotropic $\Delta n$ tensor.

\subsection{Simulation Methodology}
Three-dimensional FDTD simulations were carried out in Lumerical FDTD~\cite{taflove2005}. The simulation volume encloses the ring and both bus waveguides, bounded by perfectly matched layers on all sides. A non-uniform mesh was used, with a 20~nm step in the bulk and a 5~nm step in the waveguide cross section. The optical source was a mode-injection port at the input bus, configured for the TE or TM fundamental mode. Transmission spectra were recorded at both the through and drop ports. Silicon and silicon dioxide optical constants were taken from the Palik database~\cite{palik1998}. BeS was modeled as a diagonal anisotropic material with the DFT-derived wavelength-dependent $n$ and $k$ along each axis. Mode profiles shown in Figure~\ref{fig:modes} were computed with Lumerical MODE Solutions using the same geometry and material definitions.

\subsection{Fabrication-Tolerant Sensitivity Enhancement via Intrinsic Film Strain}
Conformal ALD growth of BeS on a silicon waveguide introduces
intrinsic biaxial compressive strain due to the lattice mismatch
between the zinc-blende BeS unit cell and the silicon substrate
surface~\cite{george2010ald}.
For typical ALD conditions this strain lies in the range of
$-2$\% to $-4$\%~\cite{nanoselect2024}.

First-principles calculations show that compressive strain in this
range increases the anisotropy of the adsorption-induced
permittivity change by enhancing the electron concentration at the
BeS surface~\cite{nanoselect2024}.
Through Equation~\eqref{eq:sensitivity2}, the sensitivity scales as
$S \propto \Gamma\,\Delta n_{\text{clad}}/n_g$.
Because strain increases $\Delta n_{\text{clad}}$ without
meaningfully altering $\Gamma$ or $n_g$, the net effect is a
proportional improvement in TM sensitivity and a corresponding
reduction in the limit of detection below the 1.5~mRIU floor
reported for the nominal geometry.

This intrinsic strain effect is advantageous rather than detrimental:
the fabrication condition that would ordinarily require careful
process control instead passively enhances sensing performance.
The 10-nm film thickness is below the critical thickness for strain
relaxation in this material system, so the strained state is stable
through the device lifetime.

\section{Performance Metrics}
Four quantities characterize the sensor response. The wavelength sensitivity,
\begin{equation}
    S = \frac{\Delta\lambda}{\Delta n_{\text{clad}}} \quad [\text{nm/RIU}],
\end{equation}
measures the resonance shift per unit change in the probed cladding index~\cite{white2008,vollmer2012}. For a guided mode with confinement factor $\Gamma$ in the cladding region and group index $n_g$, the sensitivity relates to device parameters as~\cite{bogaerts2012,chrostowski2015}
\begin{equation}
    S = \frac{\lambda_{\text{res}}\,\Gamma}{n_g}.
    \label{eq:sensitivity2}
\end{equation}
The quality factor,
\begin{equation}
    Q = \frac{\lambda_{\text{res}}}{\text{FWHM}},
\end{equation}
quantifies resonance sharpness and determines how precisely a shift can be resolved against the linewidth~\cite{vahala2003,chrostowski2015}. The figure of merit combines sensitivity and Q-factor,
\begin{equation}
    \text{FoM} = \frac{S \, Q}{\lambda_{\text{res}}} \quad [\text{RIU}^{-1}],
    \label{eq:fom}
\end{equation}
and captures the ability to resolve small index changes relative to the linewidth~\cite{white2008}. The limit of detection is evaluated assuming a minimum resolvable wavelength shift of 10~pm, which is representative of a commercial tunable-laser readout~\cite{vollmer2012,chrostowski2015},
\begin{equation}
    \text{LoD} = \frac{0.01 \; \text{nm}}{S} \quad [\text{RIU}],
    \label{eq:lod}
\end{equation}
and is reported in mRIU.

\section{Results and Discussion}
\label{sec:results}

\subsection{Transmission Spectra and Dual-Polarization Response}
\begin{figure}[htbp]
    \centering
    \includegraphics[width=\linewidth]{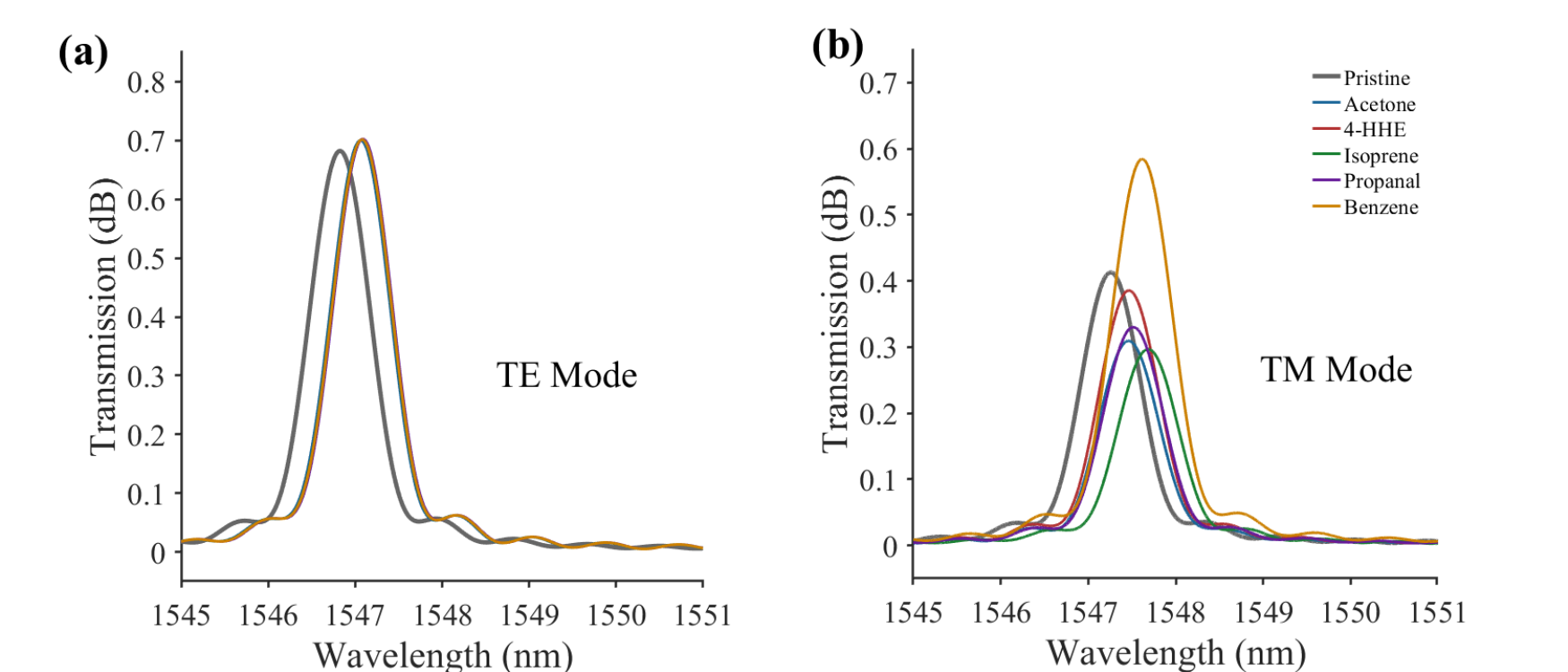}
    \caption{Drop-port transmission spectra of the BeS-clad microring near 1547~nm. (a) TE mode. All five analytes produce overlapping spectra shifted by an identical 0.263~nm from the pristine reference. (b) TM mode. Each analyte produces a distinct shift and amplitude change. Benzene uniquely shows a reduced peak, while the other four gases raise the peak amplitude.}
    \label{fig:spectra}
\end{figure}

Figure~\ref{fig:spectra} shows the drop-port transmission of the BeS-clad microring around 1547~nm for the pristine device and for the five biomarkers. Panel (a) is the TE mode and panel (b) is the TM mode. The TE spectra overlap almost exactly for all five gases. Each analyte shifts the resonance by an identical $\Delta\lambda_{TE} = 0.263$~nm from the pristine curve, with a near-uniform amplitude change near 0.02~dB. This TE degeneracy arises because the TE confinement-weighted $\Delta n_y$ produces a similar effective-index perturbation across the chosen analytes. Far from being a limitation, this behavior provides a built-in reference channel. The TE resonance reports the presence and concentration of adsorbed analyte without distinguishing chemistry. A stable reference is useful for baseline correction against temperature drift and non-specific adsorption.

The TM spectra show strong analyte-specific behavior. Each gas produces a distinct resonance shift and a distinct transmission-amplitude change. Acetone, 4-hydroxyhexenal, isoprene, and 2-propenal raise the peak transmission relative to the pristine spectrum. Benzene lowers it. The inverted response of benzene is a dispersive signature that couples the real and imaginary parts of its adsorption-induced polarizability and is not reproduced by any of the other analytes. The TM shifts span from $\Delta\lambda_{TM} = 0.200$~nm for acetone to $\Delta\lambda_{TM} = 0.426$~nm for isoprene.

\begin{figure}[htbp]
    \centering
    \includegraphics[width=\linewidth]{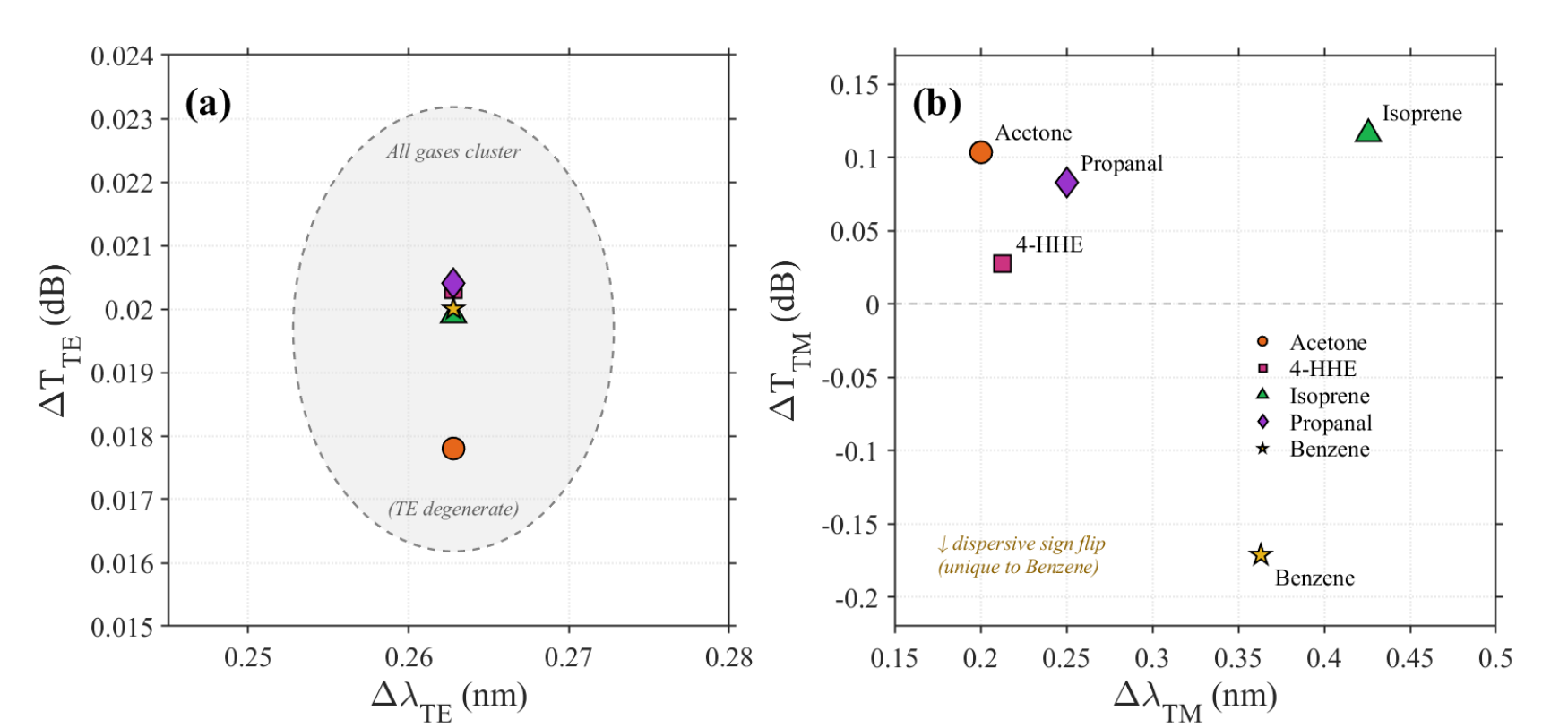}
    \caption{Device-level response metrics for the five biomarkers. (a) TE wavelength shift $\Delta\lambda_{TE}$. (b) TM wavelength shift $\Delta\lambda_{TM}$. (c) TE transmission-amplitude change $|\Delta T_{TE}|$. (d) TM transmission-amplitude change $|\Delta T_{TM}|$. The TE channel is near-degenerate across analytes, while the TM channel discriminates them.}
    \label{fig:bar_shifts}
\end{figure}

The scalar shifts and amplitude changes for both polarizations are summarized in Figure~\ref{fig:bar_shifts}. The uniform TE response and the gas-specific TM response together form the two-dimensional optical fingerprint of the device. The pair $(\Delta\lambda_{TE}, \Delta\lambda_{TM})$ alone distinguishes the five analytes. Including the amplitude changes extends the fingerprint to four dimensions and isolates benzene through its unique sign flip.

\subsection{Dielectric Origin of Polarization-Selective Shifts}
\label{sec:dielectric_origin}

The sign reversal of $\Delta\lambda_{TM}$ for \ce{CO2} and \ce{H2O}
relative to the five target biomarkers has a direct physical origin
in the surface charge redistribution that accompanies adsorption.

Hirshfeld charge-population analysis of the adsorption complexes
from first-principles calculations shows that all five target VOCs
act as net electron donors to BeS surface states~\cite{nanoselect2024}.
Acetone transfers $+0.343\,|e|$ and 4-hydroxyhexenal transfers
$+0.438\,|e|$ toward the BeS surface bonds, the two largest values
in the series.
Isoprene ($+0.258\,|e|$) and 2-propenal ($+0.482\,|e|$) follow.
This electron donation increases bond polarizability along the
out-of-plane crystallographic axis, raising $\varepsilon_{zz}$
and shifting the TM resonance to longer wavelengths
($\Delta\lambda_{TM} > 0$).

Benzene is a partial exception.
Its aromatic $\pi$ system transfers only $+0.055\,|e|$, the
smallest value in the series, consistent with weak
physisorption~\cite{nanoselect2024}.
The smaller charge transfer is accompanied by an adsorption geometry
in which the benzene ring lies flat on the BeS surface at a distance
of 2.275~\AA, compared to the 1.694--1.921~\AA\ range for the
chemisorbed species.
This geometry preferentially perturbs the in-plane bonds and
produces dispersive coupling between the real and imaginary parts
of the adsorption-induced susceptibility.
This is the physical origin of the unique sign flip in
$\Delta T_{TM}$ for benzene reported in Figure~\ref{fig:spectra}.

\ce{CO2} and \ce{H2O} interact with the BeS surface as weak
electron acceptors, with charge transfer of only $+0.004\,|e|$
for \ce{CO2}~\cite{nanoselect2024}.
Their electronegative oxygen lone pairs withdraw charge from the
out-of-plane BeS bonds, suppressing $\varepsilon_{zz}$ and
shifting the TM resonance to shorter wavelengths
($\Delta\lambda_{TM} < 0$).
The sign of $\Delta\lambda_{TM}$ therefore encodes the
electron-donor or electron-acceptor character of the adsorbate
at the atomic scale, providing a first-level chemical class filter
that requires no signal processing beyond a sign comparison.

\subsection{Sensitivity and Quality Factor}
\begin{figure}[htbp]
    \centering
    \includegraphics[width=0.95\linewidth]{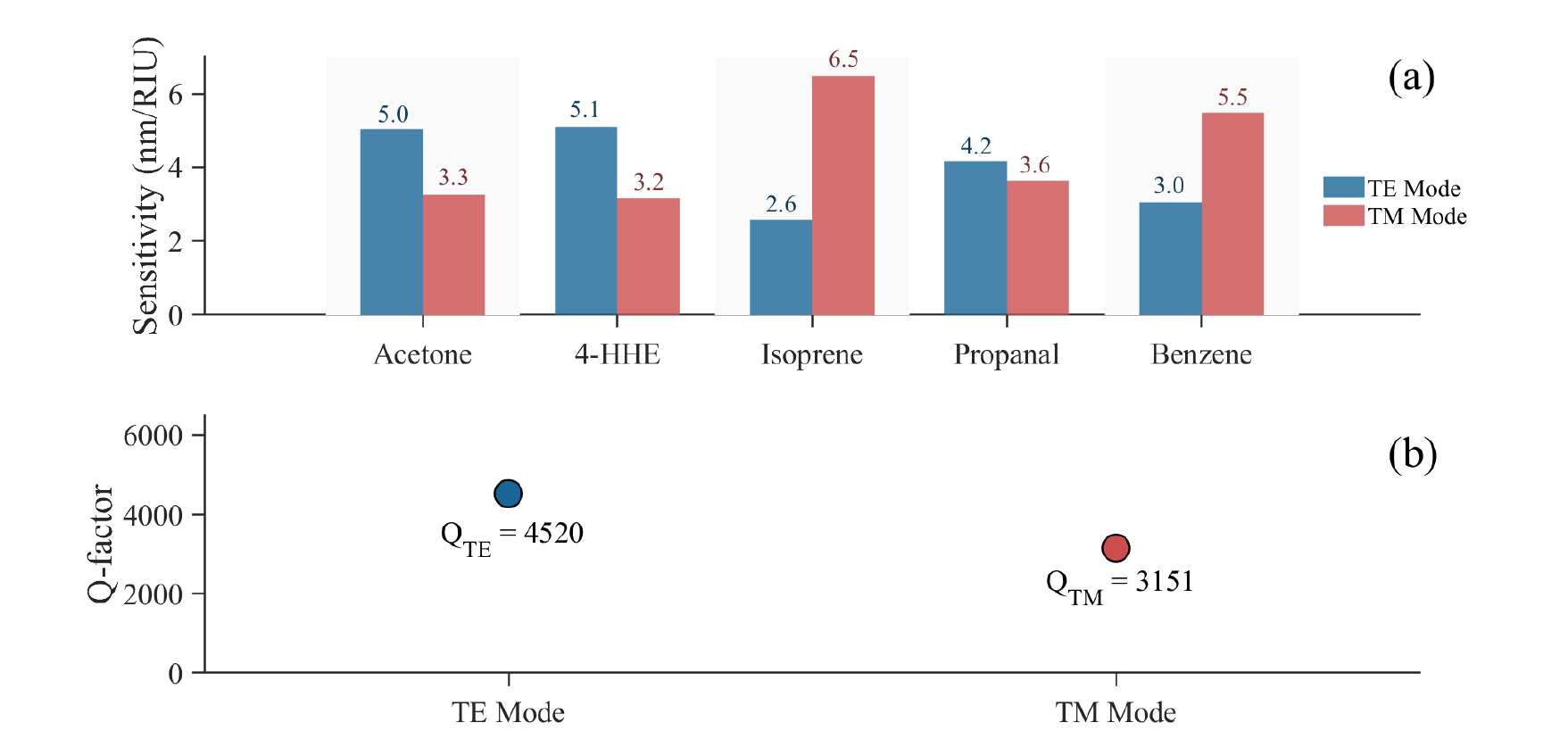}
    \caption{Mode-selective sensing performance. (a) Per-analyte sensitivity $S$ in nm/RIU for TE and TM modes. The higher-sensitivity mode differs between analytes because of the anisotropic BeS response. (b) Quality factor. $Q_{TE} = 4520$ and $Q_{TM} = 3151$. The TE mode has tighter confinement and lower radiation loss. The TM mode trades Q for stronger interaction with the cladding.}
    \label{fig:sq}
\end{figure}

The per-analyte sensitivities are compared in Figure~\ref{fig:sq}(a). The TE sensitivity ranges from 2.6~nm/RIU for isoprene to 5.1~nm/RIU for 4-hydroxyhexenal. The TM sensitivity ranges from 3.2~nm/RIU for 4-hydroxyhexenal to 6.5~nm/RIU for isoprene. The relative ordering of the two modes depends on the analyte. TE is higher for acetone and 4-hydroxyhexenal. TM is higher for isoprene, 2-propenal, and benzene. This inversion is a direct consequence of the BeS anisotropy. Gases with larger $\Delta n_y$ dominate the TE response, while gases with larger $\Delta n_z$ dominate the TM response.

Figure~\ref{fig:sq}(b) compares the quality factors. The TE mode achieves $Q_{TE} = 4520$ and the TM mode achieves $Q_{TM} = 3151$. The higher TE Q-factor reflects tighter confinement and lower bending loss. The TM mode has a larger evanescent fraction extending upward through the BeS film into the gas region, which boosts the cladding interaction at the cost of additional scattering and radiation loss. The Q-factor is therefore not a figure of merit on its own. It must be combined with sensitivity, as discussed below.

\subsection{Figure of Merit and Limit of Detection}
\begin{figure}[t]
    \centering
    \includegraphics[width=\linewidth]{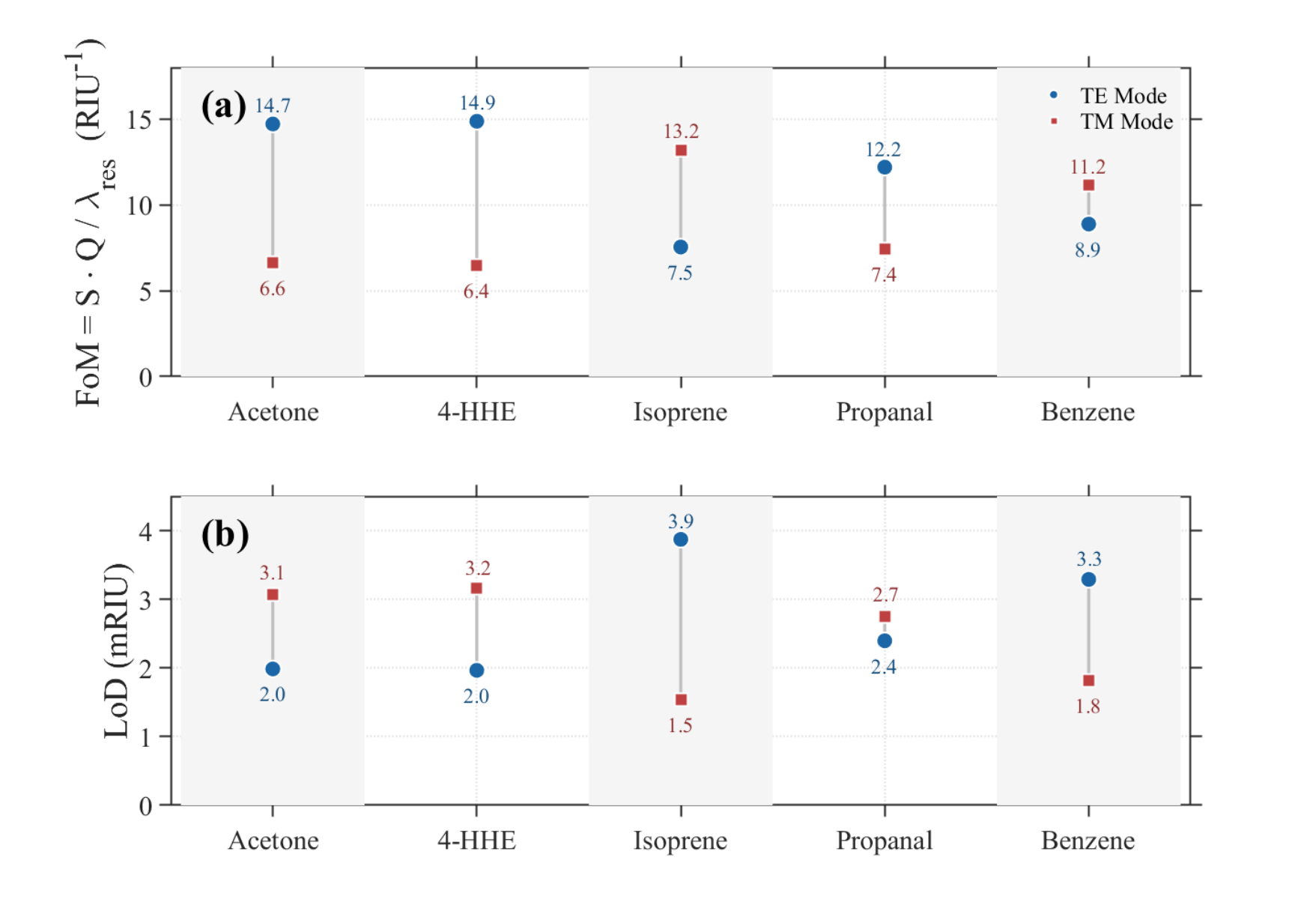}
    \caption{Figure of merit and limit of detection for the TE and TM modes. (a) $\text{FoM} = S \, Q / \lambda_{\text{res}}$. (b) $\text{LoD} = 0.01\,\text{nm}/S$. The mode that gives the best value alternates between analytes, demonstrating that TE and TM are complementary rather than competing channels.}
    \label{fig:fom_lod}
\end{figure}

Figure~\ref{fig:fom_lod} compares the two modes using the FoM and LoD metrics. The TE FoM ranges from $7.5~\text{RIU}^{-1}$ for isoprene to $14.9~\text{RIU}^{-1}$ for 4-hydroxyhexenal. The TM FoM ranges from $6.4~\text{RIU}^{-1}$ for 4-hydroxyhexenal to $13.2~\text{RIU}^{-1}$ for isoprene. The TE LoD ranges from 2.0~mRIU (acetone and 4-hydroxyhexenal) to 3.9~mRIU (isoprene). The TM LoD ranges from 1.5~mRIU (isoprene) to 3.2~mRIU (4-hydroxyhexenal). For every analyte, the mode with the smaller detection limit compensates the mode with the larger one. The two channels are complementary. Operating the same device in both polarizations therefore extracts more information per measurement than either polarization alone.

\subsection{Polarization-Selective Chemical Discrimination}
Table~\ref{tab:summary} consolidates the sensing performance of both modes for all five analytes. The central observation is that no two analytes share the same response vector when both polarizations are read. Viewed in the multivariate space spanned by $(\Delta\lambda_{TE}, \Delta\lambda_{TM}, \Delta T_{TE}, \Delta T_{TM})$, the five biomarkers occupy distinct clusters. A simple classifier with access to this vector can identify the species from a single spectral scan of a single device. This function is ordinarily provided by arrays of differently functionalized resonators~\cite{hao2023vocml}. The anisotropic BeS approach achieves the same discrimination with one physical element by exploiting directional information already carried by the cladding material.

\begin{table}[htbp]
    \centering
    \caption{Sensing performance of the BeS-clad microring for the five breath biomarkers. $\Delta n$ values are from first-principles calculations. $\Delta\lambda$ values are from FDTD simulations. Sensitivity, FoM, and LoD are evaluated at $\lambda_{\text{res}} = 1547$~nm with $Q_{TE} = 4520$ and $Q_{TM} = 3151$.}
    \label{tab:summary}
    \small
    \begin{tabular}{lcccccc}
        \toprule
        Analyte & $\Delta n_{TE}$ & $\Delta n_{TM}$ & $\Delta \lambda_{TE}$ (nm) & $\Delta \lambda_{TM}$ (nm) & $S_{TE}$ (nm/RIU) & $S_{TM}$ (nm/RIU) \\
        \midrule
        Acetone    & 0.0522 & 0.0616 & 0.263 & 0.200 & 5.0 & 3.3 \\
        4-HHE      & 0.0516 & 0.0672 & 0.263 & 0.213 & 5.1 & 3.2 \\
        Isoprene   & 0.1018 & 0.0657 & 0.263 & 0.426 & 2.6 & 6.5 \\
        2-Propenal & 0.0630 & 0.0687 & 0.263 & 0.250 & 4.2 & 3.6 \\
        Benzene    & 0.0865 & 0.0662 & 0.263 & 0.363 & 3.0 & 5.5 \\
        \bottomrule
    \end{tabular}
\end{table}

\subsection{Comparison with an Isotropic Cladding}
If the cladding were isotropic, the TE and TM modes would sample the same $\Delta n$ up to the differences in their confinement factors. The ratio $\Delta\lambda_{TE} / \Delta\lambda_{TM}$ would then be a material-independent constant fixed by the device geometry. Any deviation from this constant ratio across analytes is a direct signature of cladding anisotropy. In the present device this ratio varies from 0.62 for isoprene to 1.32 for acetone. The observed variation cannot be reproduced by any isotropic cladding, and this is what enables the chemical specificity of the dual-polarization readout.

\subsection{Cross-Sensitivity to Ambient Breath Interferents}
\begin{figure}[htbp]
    \centering
    \includegraphics[width=\linewidth]{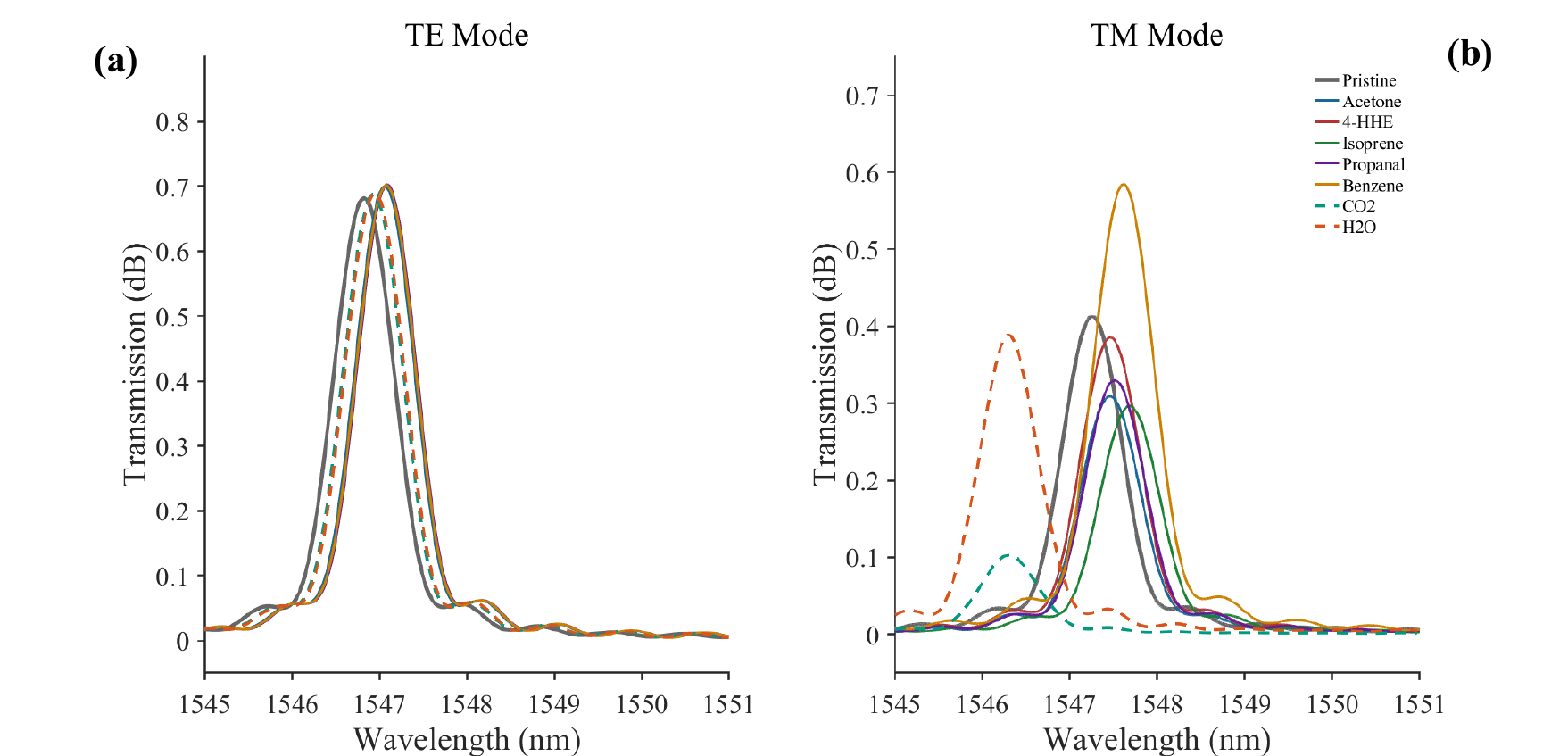}
    \caption{Transmission spectra of the BeS-clad microring for the five target biomarkers (solid lines) and the two dominant breath interferents \ce{CO2} and \ce{H2O} (dashed lines). (a) TE mode. Interferents shift the resonance by 0.104~nm (\ce{CO2}) and 0.124~nm (\ce{H2O}), both below the uniform target shift of 0.263~nm. (b) TM mode. Both interferents produce a negative resonance shift of $-0.968$~nm, placing them in the opposite direction from all five target biomarkers.}
    \label{fig:crosssens_spectra}
\end{figure}

Exhaled breath contains \ce{CO2} at approximately 4\% by volume and is saturated with water vapor~\cite{pauling1971}. Both species adsorb on the BeS surface and must be accounted for before the sensor can be considered selective. FDTD simulations were performed for \ce{CO2} and \ce{H2O} using DFT-derived anisotropic $n,k$ tensors for BeS under each adsorbate, following the same procedure as for the five target biomarkers. Figure~\ref{fig:crosssens_spectra} shows the resulting transmission spectra overlaid with the target gas responses for both polarizations.

The TM response reveals a physically important distinction. All five target VOCs produce a positive TM resonance shift, ranging from $+0.200$~nm to $+0.426$~nm. Both \ce{CO2} and \ce{H2O} produce a negative TM shift of $-0.968$~nm. The sign of $\Delta\lambda_{TM}$ alone separates interferents from target analytes with no ambiguity. This sign difference arises from the adsorption geometry: polar VOC biomarkers donate electron density to BeS surface states in a way that increases $\varepsilon_{zz}$, while the physisorption of \ce{CO2} and \ce{H2O} withdraws electron density from the out-of-plane bonds, reducing $\varepsilon_{zz}$ and shifting the TM resonance to shorter wavelengths.

\begin{figure}[t]
    \centering
    \includegraphics[width=0.85\linewidth]{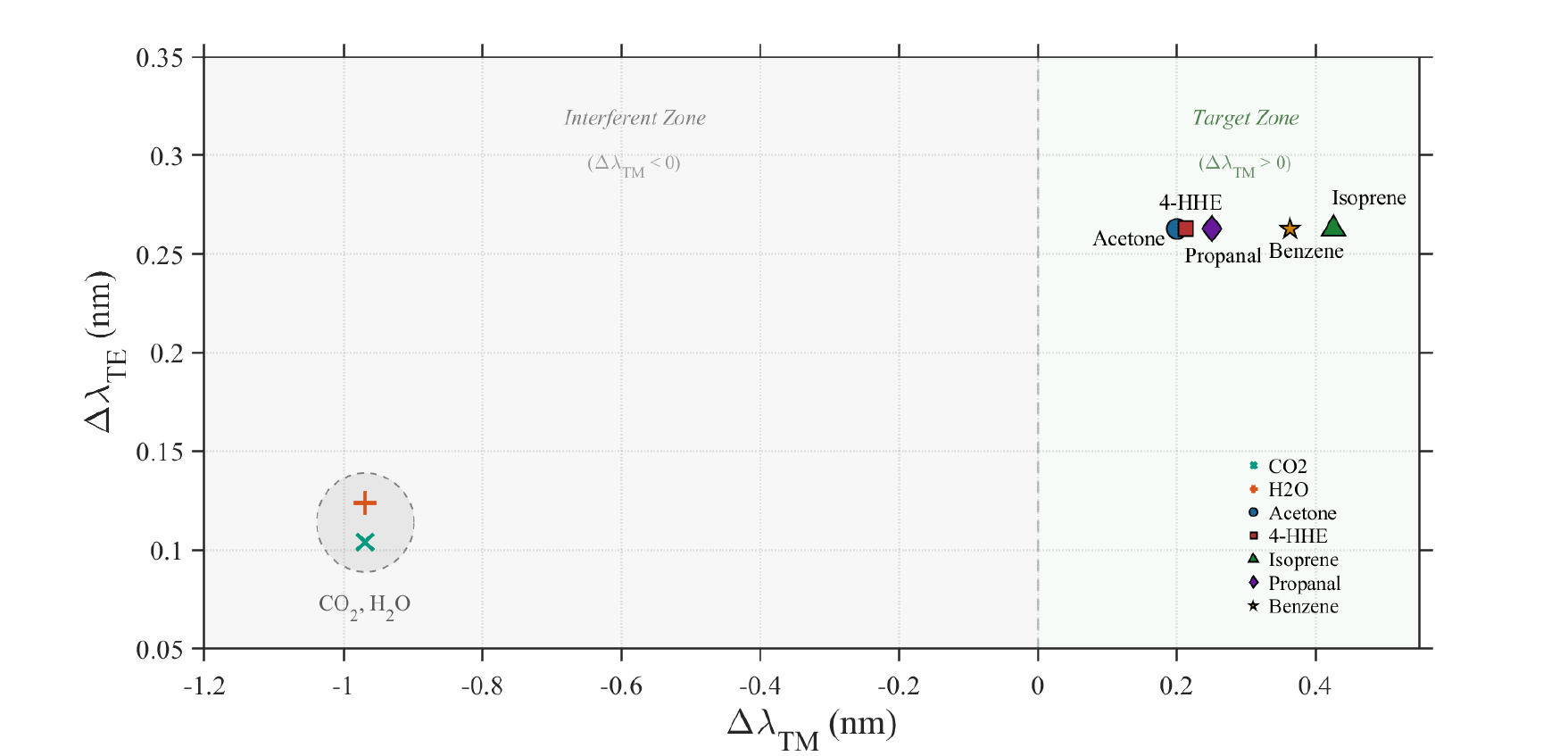}
    \caption{Updated optical fingerprint plane showing target biomarkers (filled markers) and breath interferents \ce{CO2} and \ce{H2O} (open markers). All five target biomarkers occupy the positive-$\Delta\lambda_{TM}$ half-plane. Both interferents occupy the negative-$\Delta\lambda_{TM}$ half-plane. The sign of $\Delta\lambda_{TM}$ provides a first-level discriminator that requires no additional calibration.}
    \label{fig:crosssens_fp}
\end{figure}

The TE channel provides a secondary confirmation. Target biomarkers shift the TE resonance by 0.263~nm. \ce{CO2} shifts it by 0.104~nm and \ce{H2O} by 0.124~nm, both significantly below the target level. In the $(\Delta\lambda_{TE}, \Delta\lambda_{TM})$ plane, the interferents cluster at negative $\Delta\lambda_{TM}$ while all targets cluster at positive $\Delta\lambda_{TM}$, as shown in Figure~\ref{fig:crosssens_fp}. The two populations are fully separated with no overlap.

The selectivity data are summarized in Table~\ref{tab:selectivity}. Since \ce{CO2} and \ce{H2O} both produce a TM blue shift far outside the positive range occupied by targets, a simple threshold on the sign of $\Delta\lambda_{TM}$ classifies any measurement as either a biomarker event or an interferent event before any further processing is applied. Moreover, because the TE shift from \ce{H2O} (0.124~nm) is measurable and stable, the TE channel can serve as a real-time humidity monitor. Subtracting the TE contribution of the water baseline from the total TE signal removes humidity drift from the concentration estimate of a co-present target biomarker.

\begin{table}[htbp]
    \centering
    \caption{Selectivity summary for target biomarkers and dominant breath interferents. All $\Delta\lambda$ values are from FDTD simulations referenced to the wet BeS baseline. The sign of $\Delta\lambda_{TM}$ separates all target analytes from both interferents.}
    \label{tab:selectivity}
    \small
    \begin{tabular}{lcccc}
        \toprule
        Species & Type & $\Delta\lambda_{TE}$ (nm) & $\Delta\lambda_{TM}$ (nm) & TM Sign \\
        \midrule
        Acetone    & Target      & $+0.263$ & $+0.200$ & Positive \\
        4-HHE      & Target      & $+0.263$ & $+0.213$ & Positive \\
        Isoprene   & Target      & $+0.263$ & $+0.426$ & Positive \\
        2-Propenal & Target      & $+0.263$ & $+0.250$ & Positive \\
        Benzene    & Target      & $+0.263$ & $+0.363$ & Positive \\
        \midrule
        \ce{CO2}   & Interferent & $+0.104$ & $-0.968$ & Negative \\
        \ce{H2O}   & Interferent & $+0.124$ & $-0.968$ & Negative \\
        \bottomrule
    \end{tabular}
\end{table}

These results address a key concern for simulation-only sensor studies. The BeS cladding is not universally sensitive to all adsorbates. Its anisotropic permittivity tensor responds differently to species that increase versus decrease its out-of-plane polarizability, and this difference is directly resolved by the TM channel. The sensor therefore has a built-in chemical class filter at the material level, prior to any signal processing.

\subsection{Surface Regeneration and High-Throughput Operation}

A practical breath sensor must regenerate between successive measurements.
Physisorption recovery times on BeS vary strongly between analytes
at 298~K: acetone recovers in 19.72~s, isoprene in $2.08 \times 10^{-4}$~s,
and benzene in $8.58 \times 10^{-5}$~s.
4-hydroxyhexenal presents the most demanding case with a thermal
recovery time of $1.97 \times 10^{6}$~s, which is incompatible with
clinical throughput~\cite{nanoselect2024}.

Ultraviolet photodesorption provides a practical solution.
UV illumination at photon energies at or above the BeS bandgap
($E_g > 4.5$~eV~\cite{nanoselect2024}) lowers the desorption barrier
by generating electron-hole pairs that transiently weaken the surface
dipole coupling.
Under UV irradiation at 298~K, the recovery time is reduced by three
orders of magnitude for all analytes: acetone from 19.72~s to
$1.9 \times 10^{-2}$~s, and 4-hydroxyhexenal from
$1.97 \times 10^{6}$~s to $1.97 \times 10^{3}$~s~\cite{nanoselect2024}.
For isoprene and benzene the UV-assisted times fall below
$10^{-7}$~s~\cite{nanoselect2024}, effectively instantaneous at
the measurement timescale.

A micro-UV-LED at $\lambda_{\text{UV}} < 270$~nm, mounted adjacent
to the ring at a standoff distance compatible with standard SOI
packaging, would provide this reset capability between breath samples
without perturbing the C-band optical readout.
For 4-HHE, where UV-assisted recovery remains on the order of
$10^3$~s, supplementary resistive heating to the computed desorption
temperature of 334~K provides a complementary reset path within
a clinically acceptable cycle time~\cite{nanoselect2024}.

Table~\ref{tab:master} consolidates the atomic-scale adsorption
physics from first-principles calculations~\cite{nanoselect2024}
alongside the device-level optical response from FDTD simulations,
providing a unified view of the sensing mechanism from charge
transfer at the BeS surface through to the measurable resonance
shift at the drop port.

\begin{table*}[htbp]
    \centering
    \caption{Adsorption physics and sensing response parameters for BeS--VOC complexes.}
    \label{tab:master}
    \small
    \begin{tabular}{lllccccc}
        \toprule
        Species & Type & Interaction & $Q$ & $\Delta\lambda_{TM}$ & Thermal Rec. & UV Rec. & $T_D$ \\
                &      &             & ($|e|$) & (nm) & (s) & (s) & (K) \\
        \midrule
        Acetone    & Target & Chem. & 0.343 & $+0.200$ & 19.72               & $1.9\times10^{-2}$  & 411 \\
        4-HHE      & Target & Chem. & 0.438 & $+0.213$ & $1.97\times10^{6}$  & $1.97\times10^{3}$  & 334 \\
        Isoprene   & Target & Phys. & 0.258 & $+0.426$ & $2.08\times10^{-4}$ & $2.08\times10^{-7}$ & 316 \\
        2-Propenal & Target & Chem. & 0.482 & $+0.250$ & 0.81                & $8.1\times10^{-4}$  & 310 \\
        Benzene    & Target & Phys. & 0.055 & $+0.363$ & $8.58\times10^{-5}$ & $8.58\times10^{-8}$ & 294 \\
        \midrule
        \ce{CO2}   & Interf. & Phys. & 0.004   & $-0.968$  & --- & --- & --- \\
        \ce{H2O}   & Interf. & Phys. & $<0.04$ & Baseline  & --- & --- & Baseline \\
        \bottomrule
    \end{tabular}
    \vspace{2pt}
    \begin{flushleft}
    \scriptsize
    Note: $Q$ denotes Hirshfeld charge transfer; positive values indicate electron donation
    from adsorbate to BeS~\cite{nanoselect2024}. $\Delta\lambda_{TM}$ is referenced to the
    \ce{H2O}-adsorbed baseline. Recovery times are Van't Hoff--Arrhenius estimates at
    298~K assuming a UV attempt frequency of $10^{16}$~Hz~\cite{nanoselect2024}.
    \end{flushleft}
\end{table*}

\section{Conclusion}
A polarization-selective silicon microring resonator with an anisotropic BeS cladding has been proposed for the exhaled-breath detection of five clinically relevant VOCs. The TE and TM modes probe orthogonal components of the BeS permittivity tensor and return two independent refractive-index responses per analyte from a single device. The TE mode, with $Q_{TE} = 4520$, yields a uniform 0.263~nm shift across all tested gases and acts as a concentration reference channel. The TM mode, with $Q_{TM} = 3151$, produces analyte-specific shifts between 0.200 and 0.426~nm and a dispersive amplitude signature that uniquely identifies benzene. TE sensitivities reach 5.1~nm/RIU and TM sensitivities reach 6.5~nm/RIU, with figures of merit up to $14.9~\text{RIU}^{-1}$ and detection limits as low as 1.5~mRIU. Each analyte occupies a distinct coordinate in the $(\Delta\lambda_{TE}, \Delta\lambda_{TM})$ fingerprint plane. Chemical discrimination that ordinarily requires a functionalized sensor array is achieved here through the intrinsic directional anisotropy of a single unfunctionalized cladding material.

Cross-sensitivity simulations for \ce{CO2} and \ce{H2O}, the two dominant non-target species in exhaled breath, show that both interferents produce a negative TM resonance shift of $-0.968$~nm. All five target biomarkers produce positive TM shifts. The sign of $\Delta\lambda_{TM}$ therefore provides a first-level discriminator that separates interferent events from biomarker events without additional signal processing. The TE shift from \ce{H2O} (0.124~nm) is measurable and can serve as a real-time humidity baseline reference for concentration correction.

The study is simulation-based, and several aspects require experimental confirmation. Film deposition quality, surface roughness, and waveguide-cladding interface conditions will affect both Q-factors and modal overlap in a fabricated device. The current results assume isolated single-analyte exposure, whereas real exhaled breath presents complex mixtures. Temperature cross-sensitivity is also not addressed in the present analysis and must be quantified before deployment.

Future work will pursue ALD deposition of conformal BeS films on SOI microrings, concentration-dependent sensitivity measurements, and systematic temperature cross-sensitivity characterization. Classification performance in multi-analyte breath mixtures will be evaluated. Extension to slot-waveguide geometries is of further interest, as the higher modal overlap with the cladding should increase absolute sensitivity while preserving the polarization-selective mechanism.

\begin{acknowledgement}
The authors gratefully acknowledge the financial support provided bythe Bangladesh University of Engineering and Technology (BUET) underBasic Research Grant, BUET Ref No. Sanstha/R-60/Re-9238.
\end{acknowledgement}

\bibliography{achemso-demo}

\end{document}